\documentclass{article}
\usepackage{authblk}
\usepackage{spconf,amsmath,graphicx}
\usepackage{graphicx}
\newsavebox\mysavebox
\usepackage{caption}
\usepackage{subcaption}
\usepackage{booktabs}
\usepackage{enumitem}
\usepackage [english]{babel}
\usepackage [autostyle, english = american]{csquotes}
\MakeOuterQuote{"}

\usepackage[ruled,vlined]{algorithm2e}
\usepackage{booktabs}
\usepackage[flushleft]{threeparttable}
\SetKwComment{Comment}{$\triangleright$\ }{}

\title{Improving Fairness and Robustness in  End-to-End Speech Recognition through unsupervised clustering}
\name{
\begin{tabular}{c}
Irina-Elena Veliche$^{1}$, Pascale Fung$^{1,2,3}\thanks{$^{3}$This work was done while Pascale Fung was the Distinguished Consultant for Responsible AI at Meta.}$
\end{tabular}
}
\address{$^{1}$Meta AI, USA \quad $^{2}$The Hong Kong University of Science and Technology}

\begin{document}
\ninept
\maketitle

\begin{abstract}

The challenge of fairness arises when Automatic Speech Recognition (ASR) systems do not perform equally well for all sub-groups of
the population. In the past few years there have been many improvements in overall speech recognition quality, but without any particular
focus on advancing Equality and Equity for all user groups for whom
systems do not perform well. ASR fairness is therefore also a robustness issue. Meanwhile, data privacy also takes priority in production
systems. In this paper, we present a privacy preserving approach to
improve fairness and robustness of end-to-end ASR without using metadata,
zip codes, or even speaker or utterance embeddings directly in training. We extract utterance level embeddings using a speaker ID model trained on a public dataset, which we
then use in an unsupervised fashion to create acoustic clusters. We
use cluster IDs instead of speaker utterance embeddings as extra features during model training, which shows improvements for all demographic
groups and in particular for different accents.

\end{abstract}


\section{Introduction}

With the vast performance improvement of AI systems and their pervasiveness in our daily lives, the public and regulators are demanding that the systems to be more fair and inclusive, but also more robust and privacy preserving. Automatic speech recognition (ASR) systems act as a critical interface to many AI applications such as voice assistants, speech translators, medical record transcriptions, etc. They will remain critical in future interactions between humans and machines, as well as between humans in the Metaverse. Even though the performance of ASR systems, in terms of word error rates, has approached  human level thanks to a new paradigm of modeling technique and thanks to vast amounts of collected speech data, they still do not perform equally well for all demographic or accent groups, as shown in several studies \cite{casual_conv, amazon_fairness2022, intro_cite_fairness1, intro_cite_fairness2, intro_cite_fairness3, intro_cite_fairness4, intro_cite_fairness5}. The lack of fairness/inclusivity of ASR models is a long standing problem due to the statistical nature of the training data collected from "the wild" in actual applications. Without any curating of the data, the model tends to perform well for the majority of the speakers, while not so well for speakers that are not "in the norm". One prior approach was to collect training data from a demographically and regionally balanced set of users. However, with more user awareness of their rights to data privacy, regulators have passed down strict laws that prohibit the use of demographic and other Personal \& Private Information (PPI) in building AI systems. 

In previous decades, the lack of inclusivity in ASR systems was considered as part of a robustness issue - a robust system should perform equally well for different speakers under different speaking conditions, including accent variability\cite{DBLP:conf/icassp/LiuF99}. In this paper, we focus on the fairness of ASR systems, and on improving robustness as a result of fairness. We also focus on solving this fairness and robustness issue while meeting privacy requirement. For example, in addition to metadata, such as demographic labels or zip codes, we want to avoid directly using speaker embeddings as extra input in model training, as such representations can sometimes require stricter privacy safeguards. 

In this paper, we propose clustering the training data using utterance level embeddings extracted with a speaker ID model and use the resulting cluster ID as an additional feature in training instead. At inference time, we give each utterance an "unknown" cluster ID as an additional feature. This way no speaker embeddings are used directly. The evaluation is done on two datasets where participants self-identified across demographic groups.

\section{Related Work}
There are different approaches in  fairness for speech recognition. One standard way is to balance the training dataset, such as explored in \cite{balanced_dataset1, balanced_dataset2, balanced_dataset3}. However, the challenge with this approach is that metadata with demographic categories are needed for the training data, which is often not possible due to privacy considerations. Another approach harks back to the early days of speaker-cohort based speech recognition \cite{speaker_dep_asr, speaker_dep_asr2, speaker_dep_asr3} where the system is adapted based on the test data speakers. Accented speech recognition was also first tackled using phoneme models \cite{DBLP:conf/icassp/LiuF99}.  \cite{amazon_fairness2022} is one recent work that focuses on cohort discovery and fairness mitigation. For discovery it uses ZIP code information associated with Census data and automatic cohort discovery, which leverages speaker embeddings extracted on the wake word segment, to find top and bottom cohorts. In our case, we were not allowed to use either ZIP code information or extract speaker embeddings from utterances on assistant-type data. To improve fairness it tries both oversampling of training data and modeling cohort membership in the ASR model. The last approach is essentially using a one-hot embedding as additional feature in model training, with information on cohort embedding and acoustic features. In our case this is not possible, since we are not allowed to use speaker embeddings features in training due to privacy considerations.

There is also a good amount of work on uncovering demographic disparities in speech recognition \cite{casual_conv, intro_cite_fairness1, intro_cite_fairness2, intro_cite_fairness3, intro_cite_fairness4, intro_cite_fairness5}, but they do not describe the approaches for mitigating such disparities. 

The idea of using a domain ID approach in training was first explored in \cite{domainID}, where data from different domains were put together in training, with a domain ID attached. During inference time, using the domain ID as an additional feature during training showed improvements.

\section{Methodology}
In this section we present our approach on improving fairness and robustness for ASR models, while preserving data privacy.  One important thing to note from the start is that due to privacy limitations, we are not able to use speaker metadata, speaker embeddings or zip codes as features in training or inference. This means we have no idea about the training data composition in terms of demographic or accent groups. Therefore we cannot adopt the approach of balancing the training data to improve fairness and robustness. On the other hand, we believe that acoustic data itself shows speaker and accent variability. Hence, our main idea is to cluster the acoustic data in an unsupervised fashion and use those cluster IDs as a feature in training. Given the clustering is unsupervised, we have no insight into what is in each of the resulting clusters and thereby preserving the privacy of the user data.

\subsection{Data Clustering}
As a first step, we segment the training data into 10 seconds chunks. We then extract utterance level embeddings for each of the segments, leveraging a speaker ID model trained on VoxCeleb data. VoxCeleb is a dataset comprising of more than 7k celebrity voices, most of them from en\_US. Speaker embeddings are vector representations of any given utterance. At a high level, it uses deep neural networks to transform a variable length utterance into a fixed length embedding, of dimension 192 \cite{ecapatdnn2020}. 

Using these embeddings, we then train a principal component analysis (PCA) model for dimensionality reduction and  use that to cluster the data using the K-means algorithm \cite{kmeans}. This is a method that aims to partition $n$ observations into $k$ clusters, in which each observation belongs to the cluster with the nearest mean, serving as a prototype of the cluster. For clustering we experimented with both flat or hierarchical structures. For hierarchical we first tried to cluster the data in coarser clusters (such as grouping first in 2 to 10 clusters) and then cluster those clusters again in a recursive manner, but found through experimental results that flat clustering, where we basically let the model decide how to cluster the data in a single iteration, worked the best. At the end of the clustering, each segment has an associated cluster ID. 

\subsection{Training} \label{sec_training}
\begin{figure*}
  \includegraphics[width=\textwidth]{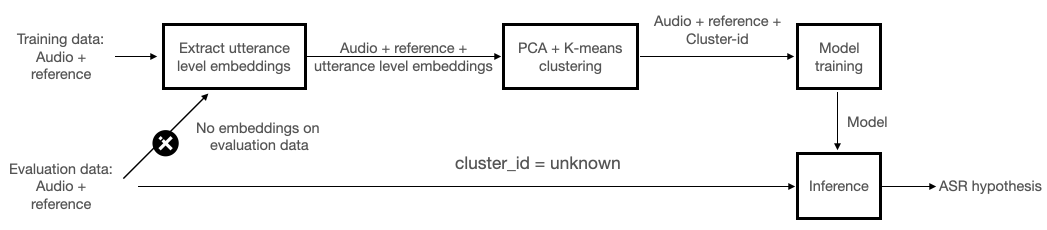}
  \caption{Speech recognition with unsupervised clustering.}
  \label{fig:flow}
\end{figure*}

Using a similar approach to  \cite{domainID}, we feed a cluster ID to the RNN-T model \cite{model_architecture} as a one-hot vector, with the ID being one of the N clusters, as described in the section above. We only feed the cluster ID to the RNN-T encoder, following the approach from \cite{multi_dialect}. An illustration  of the system flow can be seen in Figure\ref{fig:flow}.

The benefit of this multi-cluster system is that the majority of the parameters are implicitly shared by all the clusters, which forces the model to generalize across  clusters during training. We use a "masking" strategy by assigning a probability of sampling the current data into an "unknown" cluster. Unknown sampling is always into the last one-hot class. For the current experiments, the probability of sampling training examples with "unknown" speaker/cluster ID is 0.1. This ensures that during training the model will try to infer the right cluster for the unknown domain, making it more robust, even when the correct cluster ID is missing.

To determine the optimal number of clusters, we apply the elbow method \cite{elbow_method} on a subset of the training data. This consists of plotting the explained variation as a function of the number of clusters and picking the elbow of the curve as the number of clusters to use. We also validate this by experimenting with different settings during model training in terms of number of clusters in which to separate the data. Thus, we found that 50 clusters offer the optimal setup. 

\subsection{Inference}
At inference time, in order to avoid running a speaker ID model on the incoming data, which would be problematic from both a privacy and a latency perspective, we give an "unknown" cluster ID to the test data. This "unknown" cluster ID is also used in training as described in Section \ref{sec_training}. In this way we do not need to resort to a method that uses the cluster ID with the highest likelihood, as it would  slow down the inference speed. During decoding we don't use any external language model.

\section{Experimental Setup}

We trained and evaluated the approach on a publicly available data set - the Casual conversations corpus \cite{casual_conv}, as well as on an in-house dataset from the voice assistant domain. The ASR system in this investigation is an RNN-T model with an Emformer encoder \cite{model_architecture}, LSTM predictor and a joiner, having approximately 80 million parameters in total. The input feature stride is 6. Encoder network has 20 Enformer layers, each with embedded dimension of 480, 4 attention heads, feed-forward network (FFN) size 2048. Prediction and joint networks are the same as above, with 4095 word pieces. The ASR model is trained from scratch using the corresponding training utterances. \\

\subsection{Training data}
For training we used public Facebook videos in English, de-identified, that contained no personally identifiable information (PII). This is a dataset of 22K hours of manually transcribed public videos, which contain a diverse range of speakers, accents, acoustic conditions and topics.  The audio is segmented into 10 second chunks. To make the model more robust, we add additional data that is distorted using simulated reverberation and add randomly sampled additive background noise extracted from publicly available videos. Speed perturbations [32] are applied to this dataset to create two additional
training data at 0.9 and 1.1 times the original speed. We applied distortion and additive noise to the speed perturbed data, resulting in a corpus of 53.7M utterances, or 123K hours of data in total, compared to the original 22k hours. \\

\subsection{Evaluation data}
We use the following two ASR datasets for evaluation in the experiments:

\begin{itemize} [leftmargin=*]
\item \emph{Voice command:}
This is a de-identified dataset collected from a data supplier for ASR. No personally identifiable information (PII) is contained in this dataset. The participants are instructed to say voice commands on the topics of calling friends, sending a message etc. Furthermore, the speakers had the option to self-identify across demographic categories such as age, gender, ethnicity, English accent, first or home language. This allows us to  more clearly assess the gains of the new approach across demographic groups. This dataset consists of 48K utterances from 867 unique speakers. 
\item \emph{Casual conversations} \cite{casual_conv}
This is a dataset comprising of 295 hours of transcribed speech (the evaluation part), with metadata attached for age, gender and skin tone. Since skin tone does not correlate directly with speech recognition performance, as it is a purely visual characteristic, we choose to use it for illustration purposes only and not report results categorized by skin tone.
\end{itemize}

\section{Results} \label{sec_results}
\graphicspath{ {./} }
\subsection{Effect of clustering}
In order to gain insights into the clustering and only for illustration purposes, we studied the clustering of a set of evaluation data utterances and colored the data points according to linguistic, speaker demographic information and geographic information. The colors in the  clustering are used for illustration only. In both training and inference, we do not use any of these speaker or demographic information. In each of these clustering examples, we vary a single demographic characteristic at a time, to isolate and observe its correlation with acoustic clustering. For example, if we look at geographic variation, we keep gender and ethnicity fixed. All the demographic information is self-identified.

\begin{figure*}[h!]
  \begin{subfigure}[b]{0.24\textwidth}
    \includegraphics[width=\textwidth]{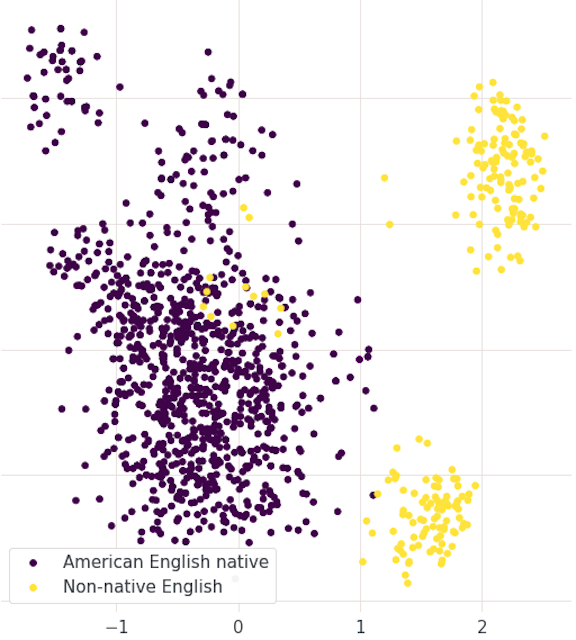}
    \caption{Native non-American English (British, Canadian, Australian, New Zealand, etc.) vs. Non-native English utterances are distinct.}
    \label{fig:f1}
  \end{subfigure}
  \hfill
  \begin{subfigure}[b]{0.24\textwidth}
    \includegraphics[width=\textwidth]{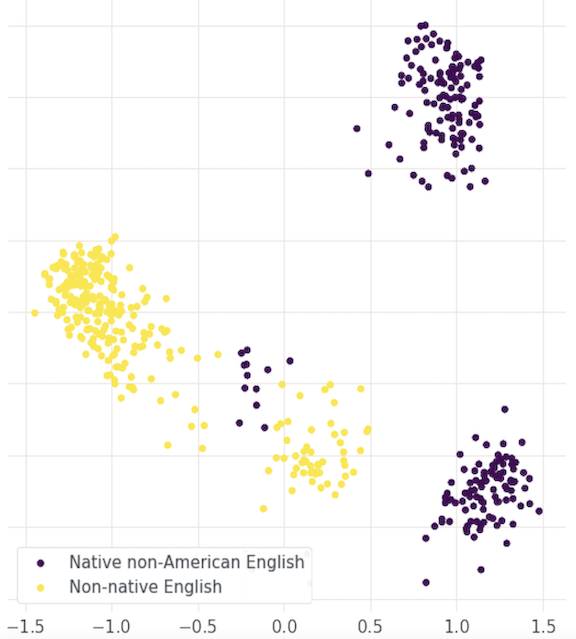}
    \caption{Utterances from American English native speakers vs. non-native speakers are distinct.}
    \label{fig:f2}
  \end{subfigure}
  \hfill
   \begin{subfigure}[b]{0.24\textwidth}
    \includegraphics[width=\textwidth]{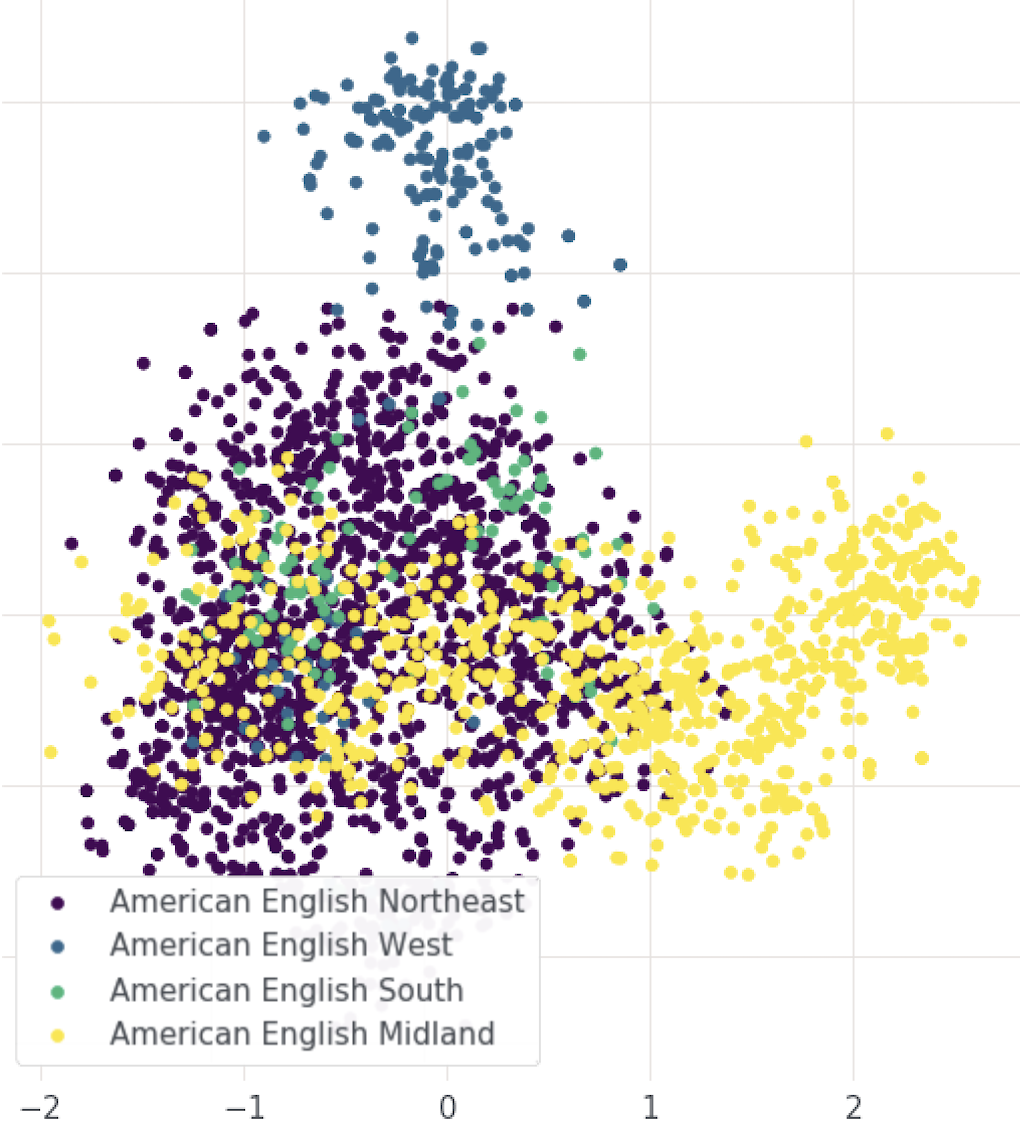}
    \caption{Utterances of American English native speakers living in Northeast, West, South, and Midland U.S. are not distinct.}
    \label{fig:f3}
  \end{subfigure}
  \hfill
  \begin{subfigure}[b]{0.24\textwidth}
    \includegraphics[width=\textwidth]{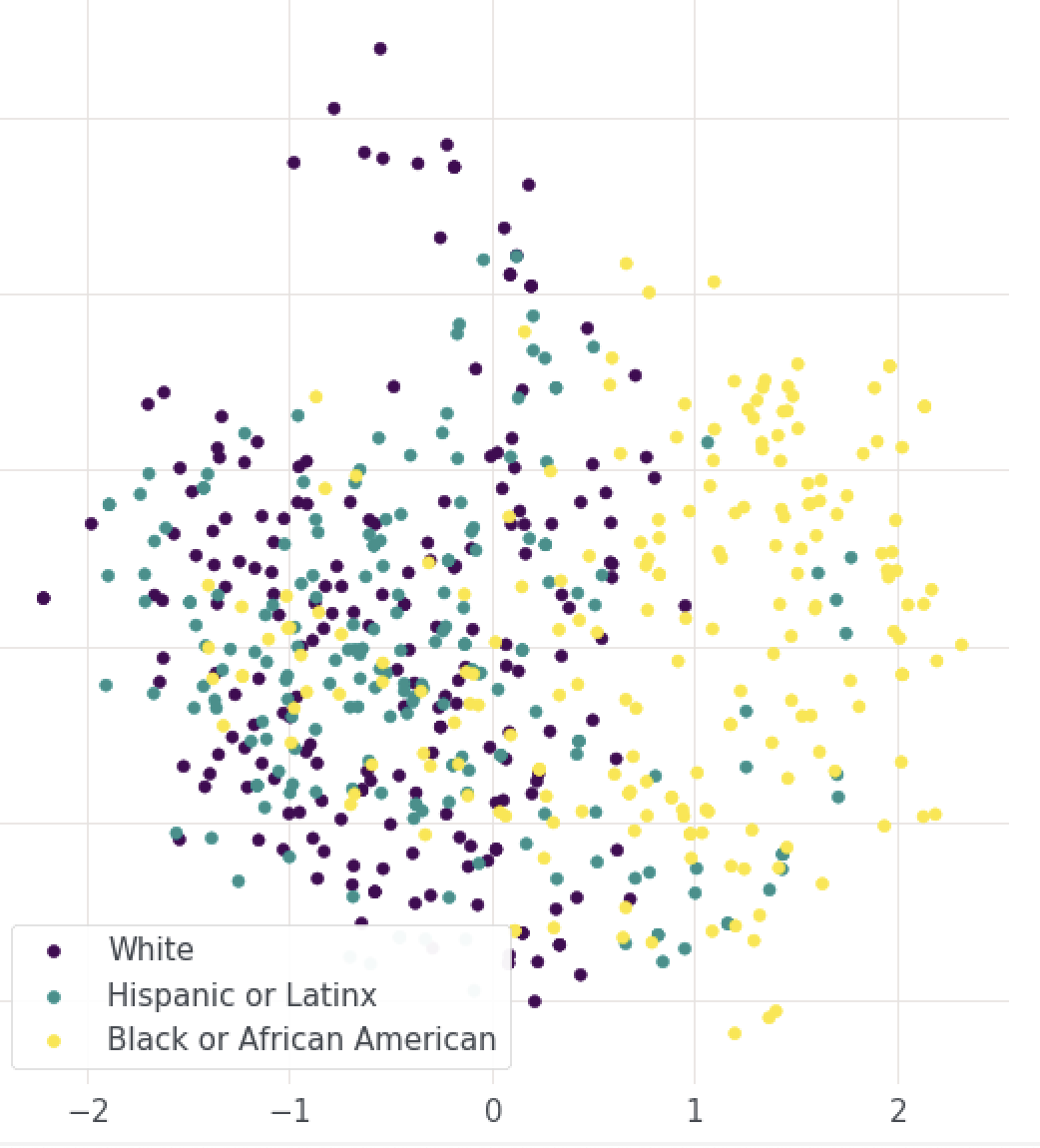}
    \caption{Utterances from black, Hispanic/Latinx and white speakers living in  Northeast U.S.}
    \label{fig:f4}
  \end{subfigure}
  \hfill
  \begin{subfigure}[b]{0.24\textwidth}
    \includegraphics[width=\textwidth]{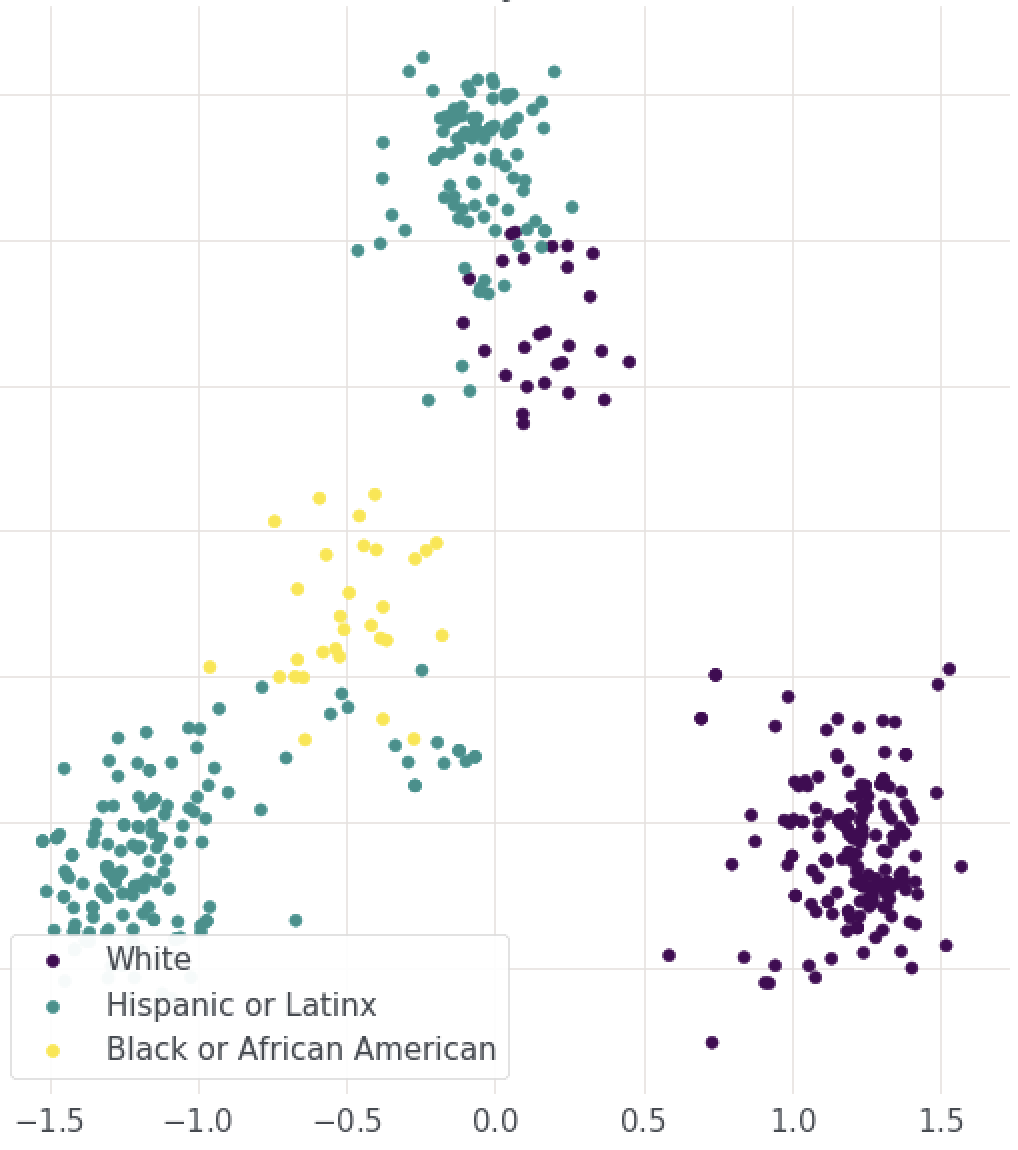}
    \caption{Utterances from black, Hispanic/Latinx and white speakers living in Western U.S.}
    \label{fig:f5}
  \end{subfigure}
  \hfill
  \begin{subfigure}[b]{0.24\textwidth}
    \includegraphics[width=\textwidth]{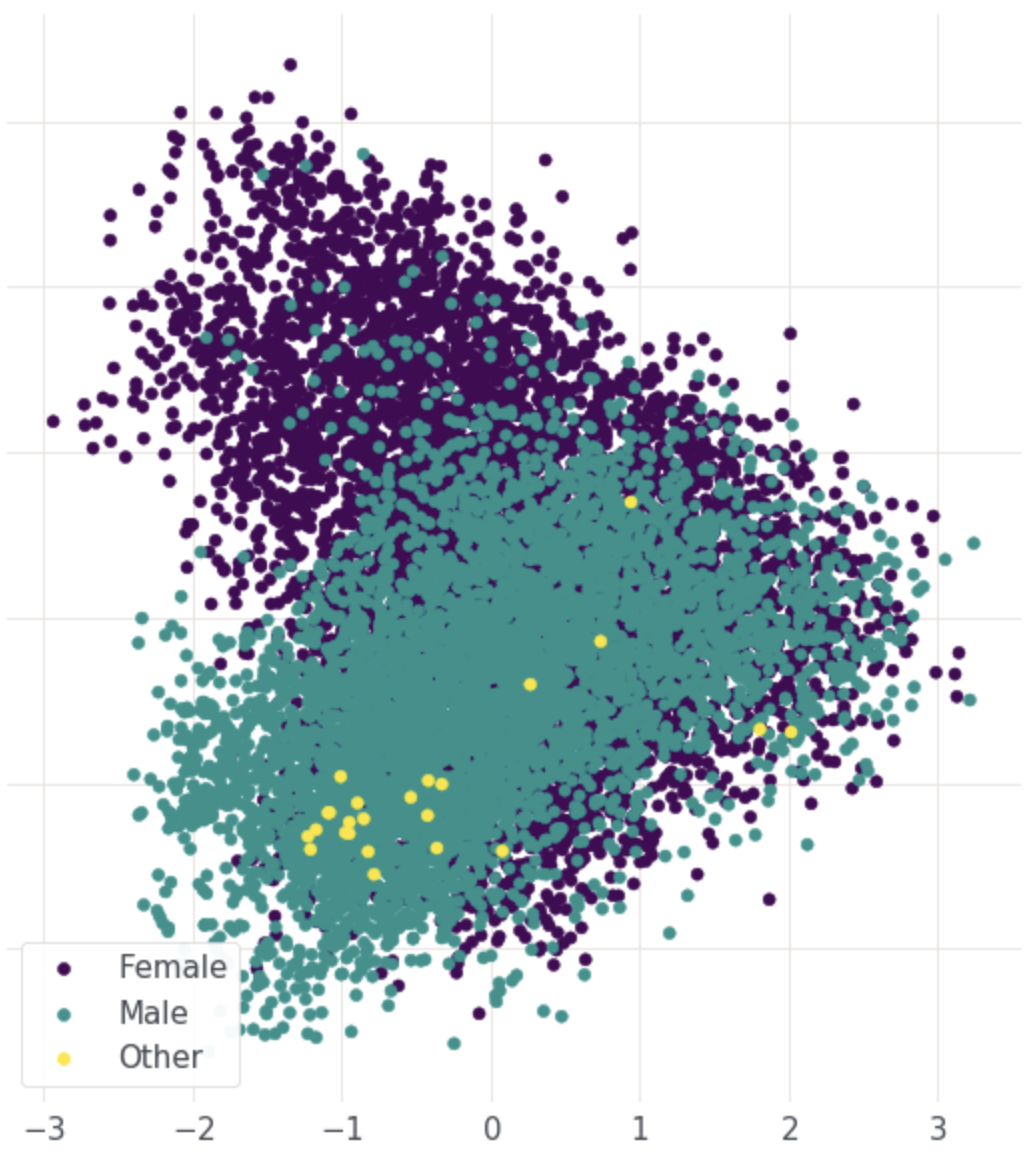}
    \caption{Utterances from different self-identified gender groups.}
    \label{fig:f6}
  \end{subfigure}
  \hfill
  \begin{subfigure}[b]{0.24\textwidth}
    \includegraphics[width=\textwidth]{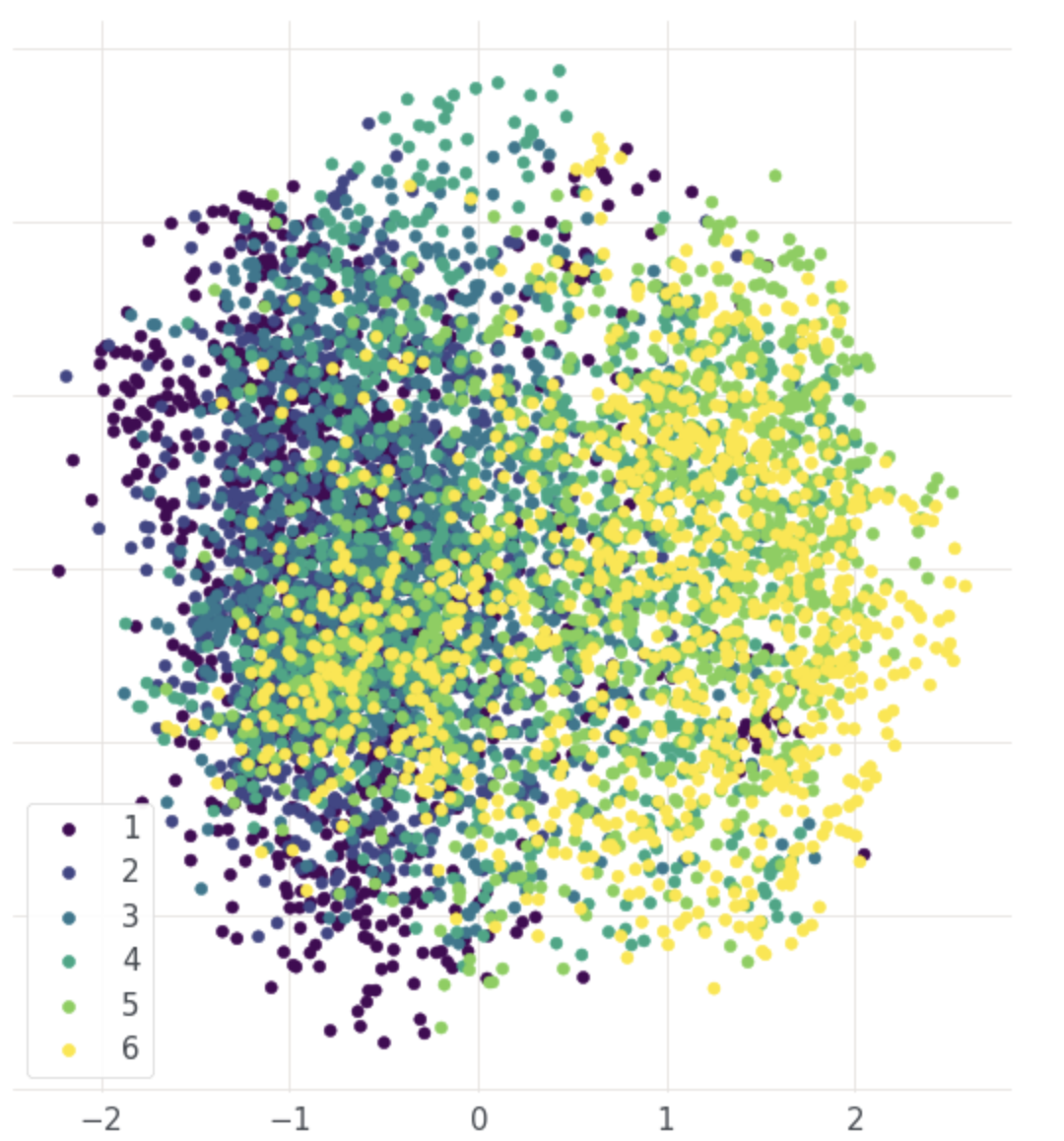}
    \caption{Utterances from  annotated skin type groups are not acoustically distinct.}
    \label{fig:f7}
  \end{subfigure}
  \hfill
  \begin{subfigure}[b]{0.24\textwidth}
    \includegraphics[width=\textwidth]{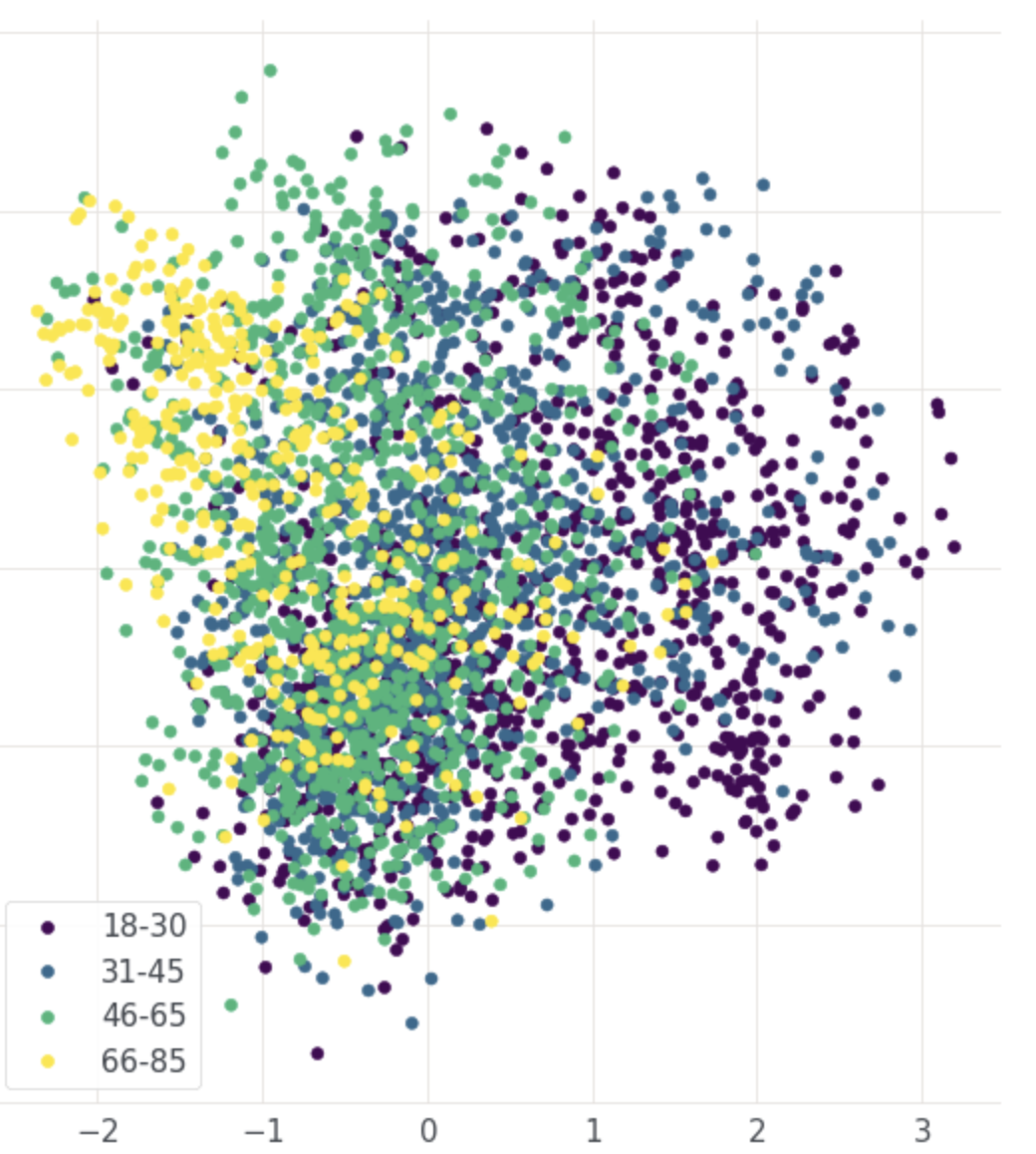}
    \caption{Utterances from different age groups of female speakers.}
    \label{fig:f8}
  \end{subfigure}
  \caption{Utterance cluster examples labeled with self-identified demographic  and geographic information.}
  \label{fig:clusters}
\end{figure*}

Figures~\ref{fig:f1} and \ref{fig:f2} show that data from different linguistic accent groups are acoustically distinct and thus tend to cluster together. For example, Figure~\ref{fig:f1} shows utterances from non-native American accent and native non-American English accent (British, Australian, New Zealand etc). When we look at different accents within the US, such as in Figure\ref{fig:f3}, the clustering is not so clear anymore, which means that  geographic labels are not necessarily good proxies for accent labels. We reason that people living in the same geographic region can have different accents if they came from other parts of the country.  This also can mean that zip codes would not be suitable proxies for accent. 

We also looked at demographic clustering, using the Casual Conversations dataset. Again, the clustering is not always clear. Figure\ref{fig:f7} shows that skin tone is not a good indicator of accent. Age groups, shown in Figure\ref{fig:f8} do not yield clear clusters, but data from the same age group  tend to be closer to each other, based on which age group it belongs in. 

The most acoustically distinct groups are according to (self-identified) gender divisions, shown in Figure\ref{fig:f6}. This makes sense as a majority of male and female speakers have different vocal cord and vocal tract characteristics. Ethnicity differences (Figures\ref{fig:f4}, \ref{fig:f5}) might lead to distinct accents if the speakers lived in the same region. Even then, Hispanic/Latinx speakers tend to cluster together with either white or black speakers. 

Generally, what we see is that geographic and self-identified ethnicity labels are bad indicators of accent or speaker variability. Therefore, we suggest that unsupervised clustering is better than using metadata for fairness and robustness of the system. \\

\subsection{Demographic group evaluation}
First, we evaluated on the voice command dataset described above. The results are shown in  Table~\ref{tab:fairness_eval}. The cluster-based approach with "unknown" cluster ID during decoding improves the overall performance of the model, across all demographic groups. However, by far the largest gains are with respect to accent, as in the groups of  non-American English native speakers and for native speakers from  Southern US, as well as for people who have Spanish as their home language or identify as Hispanic or Latinx. We also looked at whether speakers might use expressions unique to their categories, but didn't find an indication for this, meaning that the cluster ID most likely influences the encoder more than the decoder. We also note that there is a performance improvement on all demographic groups including the majority group. This is a strong indication that this approach not only improves the fairness of the model, but also its robustness.

For research purpose, we experimented with using the correct cluster ID during decoding, to have a sense of the gap. It showed improvements of 5\% rel.WER on average across demographic groups, compared to using the "unknown" cluster ID in decoding.

\begin{table}
\large
 \caption{Evaluation results on the voice command dataset.}
 \label{tab:fairness_eval}
  \centering
  \resizebox{\columnwidth}{!}{%
  \begin{tabular}{l|c|c|c|c}
    \toprule
    \cmidrule(r){2-5}    
    \emph{English accent}& \shortstack{\emph{Baseline WER\%}} & \shortstack{\emph{Cluster-based WER\%}} &  \shortstack{\emph{Relative difference}} & \shortstack{\emph{Number of Utterances}} \\
    \midrule
    \emph{Native Northeast US} & 7.25 & 6.9 & 4.82\% & 32533 \\
    \emph{Native West US} & 6.7 & 6.31 & 5.82\% & 1361\\
    \emph{Native Midland US} & 4.72 & 4.64 & 1.69\% & 2088 \\
    \emph{Native South US} & 7.42 & 6.86 & 7.54\% & 1159\\
    \emph{Native non-US} & 11.16 & 10.36 & 7.16\% & 1839\\
    \emph{Non-native} & 8.52 & 8.44 & 0.93\% & 5799\\
    \bottomrule
    \toprule
    \cmidrule(r){2-5}    
    \emph{Ethnicity}& \shortstack{\emph{Baseline WER\%}} & \shortstack{\emph{Cluster-based WER\%}} &  \shortstack{\emph{Relative difference}} & \shortstack{\emph{Number of Utterances}} \\
    \midrule
    \emph{Black or African American} & 8.65 & 8.31 & 3.93\% & 12293\\
    \emph{Hispanic or Latinx} & 7.5 & 7.19 & 4.13\% & 11690 \\
    \emph{White} & 6.15 & 5.79 & 5.85\% & 15208\\
    \emph{East Asian} & 6.39 & 5.89 & 7.82\% & 2438\\
    \emph{Southeast Asian} & 8.7 & 8.34 & 4.13\% & 1536\\
    \emph{South Asian} & 8.66 & 8.62 & 0.04\% & 1238\\
    \bottomrule
    \toprule
    \cmidrule(r){2-5}    
    \emph{Home Language}& \shortstack{\emph{Baseline WER\%}} & \shortstack{\emph{Cluster-based WER\%}} &  \shortstack{\emph{Relative difference}} & \shortstack{\emph{Number of Utterances}} \\
    \midrule
    \emph{English} & 7.23 & 6.88 & 4.84\% & 38754 \\
    \emph{Spanish} & 9.28 & 8.77 & 5.49\% & 1508 \\
    \bottomrule
    \toprule
    \cmidrule(r){2-5}    
    \emph{Gender}& \shortstack{\emph{Baseline WER\%}} & \shortstack{\emph{Cluster-based WER\%}} & \shortstack{\emph{Relative difference}} & \shortstack{\emph{Number of Utterances}} \\
    \midrule
    \emph{Female} & 6.42 & 6.02 & 6.37\% & 23646\\
    \emph{Male} & 8.45 & 8.21 & 2.84\% & 23859 \\
    \emph{Other} & 4.9 & 4.08 & 16.73\% & 181\\
    \bottomrule
  \end{tabular}
  }
  \label{tab:data}
\end{table}

The results of the evaluation on the Casual Conversations dataset are shown in Table~\ref{tab:casual_conv}. This dataset does not  have any accent information. However, as mentioned above, we used age and gender as characteristics that have impact either directly or indirectly on the speech recognition performance.

First, the cluster-based model improves performance on all age groups overall, with the largest gain in the elderly group of 66-85. Usually this group is also under-represented in the training data, so this improvement is a significant win in terms of demographic inclusivity. 

When looking at gender groups, we see a small, statistically insignificant improvement on the female subgroup. However, female speakers are the majority group for which the system already had much better performance than for the other groups. This is in line with the benchmarks from \cite{casual_conv}. For male speakers, there is a statistically significant gain of 11.17\% WER using the cluster-based approach. For the "other" gender group, the dataset is too small and the results are not statistically significant.

\begin{table}
 \caption{Evaluation results on Casual Conversations dataset.}
 \label{tab:casual_conv}
  \centering
  \resizebox{\columnwidth}{!}{%
  \begin{tabular}{l|c|c|c|c}
    \toprule
    \cmidrule(r){2-5}    
    \emph{Gender}& \shortstack{\emph{Baseline WER\%}} & \shortstack{\emph{Cluster-based WER\%}} & \shortstack{\emph{Relative difference}} & \shortstack{\emph{Number of Utterances}} \\
    \midrule
    \emph{Female} & 10.94 & 10.75 & 1.73\% & 19497 \\
    \emph{Male} & 17.54 & 15.58 & 11.17\% & 15661 \\
    \emph{Other} & 17.25 & 15.25 & 11.59\% & 768 \\
    \bottomrule
    \toprule
    \cmidrule(r){2-5}    
    \emph{Age}& \shortstack{\emph{Baseline WER\%}} & \shortstack{\emph{Cluster-based WER\%}} & \shortstack{\emph{Relative difference}} & \shortstack{\emph{Number of Utterances}} \\
    \midrule
    \emph{18-30} & 14.16 & 13.21 & 6.7\% & 12684 \\
    \emph{31-45} & 13.85 & 12.7 & 8.3\% & 11104 \\
    \emph{46-65} & 13.32 & 12.55 & 5.78\% & 9576 \\
    \emph{66-85} & 15.03 & 13.83 & 7.98\% & 1774 \\
    \bottomrule
  \end{tabular}
  }
  \label{tab:data}
\end{table}

\section{Conclusions}
In this paper, we introduced a new technique to improve the fairness and robustness of ASR models. We extract utterance embeddings from the training data using a speaker ID model, then use unsupervised clustering on these embeddings to obtain a cluster ID for each utterance, which we use as an extra feature in training. At inference time, each utterance is assigned an "unknown" cluster ID as an additional feature. Experimental results show that this approach improves model performance on all  demographics groups. Our approach also has the advantage of preserving data privacy by not using utterance embeddings directly. In addition, our approach performs ASR decoding without explicitly searching for the cluster ID, thereby saving processing time.   

\section{Acknowledgments}
The authors of this paper would like to give special thanks to Jonathan Shaw, for the speaker ID model, Yuan Shangguan for the "masking"  implementation and Duc Le for helpful discussions.


\clearpage
\bibliographystyle{IEEEbib}
\bibliography{strings,refs}

\begin{thebibliography}{10}

\bibitem{casual_conv}
Chunxi Liu, Michael Picheny, Leda Sarı, Pooja Chitkara, Alex Xiao, Xiaohui
  Zhang, Mark Chou, Andres Alvarado, Caner Hazirbas, and Yatharth Saraf,
\newblock ``Towards measuring fairness in speech recognition: Casual
  conversations dataset transcriptions,''
\newblock in {\em ICASSP 2022 - 2022 IEEE International Conference on
  Acoustics, Speech and Signal Processing (ICASSP)}, 2022, pp. 6162--6166.

\bibitem{amazon_fairness2022}
Pranav Dheram, Murugesan Ramakrishnan, Anirudh Raju, I-FAN CHEN, Brian King,
  Katherine Powell, Melissa Saboowala, Karan Shetty, and Andreas Stolcke,
\newblock ``Toward fairness in speech recognition: Discovery and mitigation of
  performance disparities,''
\newblock in {\em Interspeech 2022}, 2022.

\bibitem{intro_cite_fairness1}
Allison Koenecke, Andrew Nam, Emily Lake, Joe Nudell, Minnie Quartey, Zion
  Mengesha, Connor Toups, John Rickford, Dan Jurafsky, and Sharad Goel,
\newblock ``Racial disparities in automated speech recognition,''
\newblock {\em Proceedings of the National Academy of Sciences}, vol. 117, pp.
  201915768, 03 2020.

\bibitem{intro_cite_fairness2}
Z.~Mengesha C.~Heldreth, M.~Lahav, J.~Sublewski, and E.~Tuennerman,
\newblock ``"i don’t think these devices are very culturally sensitive." -
  the impact of errors on african americans in automated speech recognition,''
\newblock 2021.

\bibitem{intro_cite_fairness3}
Siyuan Feng, Olya Kudina, Bence Halpern, and Odette Scharenborg,
\newblock ``Quantifying bias in automatic speech recognition,'' 03 2021.

\bibitem{intro_cite_fairness4}
Morgane Riviere, Jade Copet, and Gabriel Synnaeve,
\newblock ``Asr4real: An extended benchmark for speech models,'' 10 2021.

\bibitem{intro_cite_fairness5}
J.~P. Bajorek,
\newblock ``A voice recognition still has significant race and gender biases,''
  5 2019.

\bibitem{DBLP:conf/icassp/LiuF99}
Wai~Kat Liu and Pascale Fung,
\newblock ``Fast accent identification and accented speech recognition,''
\newblock in {\em Proceedings of the 1999 {IEEE} International Conference on
  Acoustics, Speech, and Signal Processing, {ICASSP} '99, Phoenix, Arizona,
  USA, March 15-19, 1999}. 1999, pp. 221--224, {IEEE} Computer Society.

\bibitem{balanced_dataset1}
R.~Anand, K.G. Mehrotra, C.K. Mohan, and S.~Ranka,
\newblock ``An improved algorithm for neural network classification of
  imbalanced training sets,''
\newblock {\em IEEE Transactions on Neural Networks}, vol. 4, no. 6, pp.
  962--969, 1993.

\bibitem{balanced_dataset2}
Nathalie Japkowicz and Shaju Stephen,
\newblock ``The class imbalance problem: A systematic study,''
\newblock {\em Intelligent Data Analysis}, pp. 429--449, 2002.

\bibitem{balanced_dataset3}
Bartosz Krawczyk,
\newblock ``Learning from imbalanced data: Open challenges and future
  directions,''
\newblock {\em Progress in Artificial Intelligence}, vol. 5, 04 2016.

\bibitem{speaker_dep_asr}
M.~Padmanabhan, L.R. Bahl, D.~Nahamoo, and M.A. Picheny,
\newblock ``Speaker clustering and transformation for speaker adaptation in
  large-vocabulary speech recognition systems,''
\newblock in {\em 1996 IEEE International Conference on Acoustics, Speech, and
  Signal Processing Conference Proceedings}, 1996, vol.~2, pp. 701--704 vol. 2.

\bibitem{speaker_dep_asr2}
L.~Mathan and L.~Miclet,
\newblock ``Speaker hierarchical clustering for improving speaker-independent
  hmm word recognition,''
\newblock in {\em International Conference on Acoustics, Speech, and Signal
  Processing}, 1990, pp. 149--152 vol.1.

\bibitem{speaker_dep_asr3}
D.A. Reynolds and L.P. Heck,
\newblock ``Integration of speaker and speech recognition systems,''
\newblock in {\em [Proceedings] ICASSP 91: 1991 International Conference on
  Acoustics, Speech, and Signal Processing}, 1991, pp. 869--872 vol.2.

\bibitem{domainID}
Tara Sainath, Yanzhang He, Bo~Li, Arun Narayanan, Ruoming Pang, Antoine
  Bruguier, Shuo-yiin Chang, Wei Li, Raziel Alvarez, Zhifeng Chen, Chung-Cheng
  Chiu, David Garcia, Alex Gruenstein, Ke~Hu, Minho Jin, Anjuli Kannan, Qiao
  Liang, Ian McGraw, Cal Peyser, and Ding Zhao,
\newblock ``A streaming on-device end-to-end model surpassing server-side
  conventional model quality and latency,'' 03 2020.

\bibitem{ecapatdnn2020}
Brecht Desplanques, Jenthe Thienpondt, and Kris Demuynck,
\newblock ``Ecapa-tdnn: Emphasized channel attention, propagation and
  aggregation in tdnn based speaker verification,''
\newblock in {\em INTERSPEECH}, 2020, pp. 3830--3834.

\bibitem{kmeans}
Stuart~P. Lloyd,
\newblock ``Eleast squares quantization in pcm.,''
\newblock in {\em IEEE Transactions}, 1982, pp. 129--137.

\bibitem{model_architecture}
Yangyang Shi, Yongqiang Wang, Chunyang Wu, Ching-Feng Yeh, Julian Chan, Frank
  Zhang, Duc Le, and Mike Seltzer,
\newblock ``Emformer: Efficient memory transformer based acoustic model for low
  latency streaming speech recognition,''
\newblock in {\em ICASSP 2021 - 2021 IEEE International Conference on
  Acoustics, Speech and Signal Processing (ICASSP)}, 2021, pp. 6783--6787.

\bibitem{multi_dialect}
Bo~Li, Tara~N. Sainath, Khe~Chai Sim, Michiel Bacchiani, Eugene Weinstein,
  Patrick Nguyen, Zhifeng Chen, Yanghui Wu, and Kanishka Rao,
\newblock ``Multi-dialect speech recognition with a single sequence-to-sequence
  model,''
\newblock in {\em 2018 IEEE International Conference on Acoustics, Speech and
  Signal Processing (ICASSP)}, 2018, pp. 4749--4753.

\bibitem{elbow_method}
Chunhui Yuan and Haitao Yang,
\newblock ``Research on k-value selection method of k-means clustering
  algorithm,''
\newblock {\em J}, vol. 2, pp. 226--235, 06 2019.

\end{thebibliography}

\end{document}